\documentclass[epj,nopacs]{svjour}
\usepackage{graphicx}
\usepackage{bm}
\usepackage{url}
\usepackage[dvipsnames]{xcolor}
\usepackage{amsmath}
\usepackage{amssymb}
\usepackage{hyperref}
\usepackage{cite}

\def\cevns{{CE$\nu$NS}}
\def\ve{{$\bar{\nu}_{e} -e^{-}$}}

\begin{document}
\title{Cryogenic pure CsI as a probe for neutrino electromagnetic interactions}
\author{C.M.\ Lewis\inst{1,2} 
\thanks{marklewis@uchicago.edu , mark.lewis@dipc.org}
}                     
%
%
\institute{Enrico Fermi Institute, Kavli Institute for Cosmological Physics, and Department of Physics\\
University of Chicago, Chicago, Illinois 60637, USA \and Donostia International Physics Center (DIPC)\\
Paseo Manuel Lardizabal 4, 20018 Donostia-San Sebastian, Spain}

%

\date{\today}
%
\abstract{
Searches for neutrino electromagnetic interactions at reactor sites require an unusual combination of ultra-low thresholds and a stable low-background environment. It is shown here that cryogenic undoped cesium iodide (CsI) naturally satisfies these conditions in a way prior detectors have not.
Although suppression of nuclear recoil ionization efficiency at low energies limits the use of this scintillator for coherent elastic neutrino-nucleus scattering, that same property renders the detector effectively blind to those nuclear recoils from MeV-scale reactor antineutrinos.
This leaves the low-energy regime free to expose neutrino-electron (\ve) scattering as the dominant observable channel and converts cryogenic CsI into a targeted probe of electromagnetic couplings.
This work presents a conceptual design based on pure CsI crystals immersed in an active xenon-doped liquid argon veto evaluated under realistic intrinsic and environmental backgrounds. Under present detector capabilities, order-of-magnitude improvements over current reactor limits on the neutrino magnetic moment and millicharge are achievable.
Cryogenic pure CsI therefore offers a distinctive and scalable route to leading studies of \ve\ physics.
} 
\maketitle

\section{Introduction}

The viability of pure (i.e. undoped) CsI as a cryogenic medium for the detection of neutrino-nucleus ($\nu_{N} -N$) scattering, particularly coherent elastic neutrino-nucleus scattering (\cevns), has been acknowledged in recent years by a variety of groups \cite{ESS,csiqf,clovers,kims_csi_2025,chireactor,cohcsi,jparc_pheno}.
Its appeal is clear: the high atomic numbers of Cs and I enhance the \cevns\ rate, while cryogenic operation significantly increases scintillation yield, lowering the detectable energy threshold.
However, recent measurements of the quenching factor (QF, \cite{csiqf,csiqf_lowE,lewis,prior}), recently down to sub-keV$_{nr}$, show that the ionization efficiency of nuclear recoils falls sufficiently rapidly at low recoil energies that its apparent advantages for reactor \cevns\ are largely lost.
The $\sim$3 keV nuclear recoil thresholds then achievable, demonstrated in \cite{csiqf_lowE,ESS,lewis,chireactor}, are not so gainful over the $\sim$5 keV energy thresholds achieved by the first \cevns\ experiments with this scintillator family \cite{science,NIMcenns,nicole,bjorn}.

\begin{figure}
    \centering
    \includegraphics[width=0.99\linewidth]{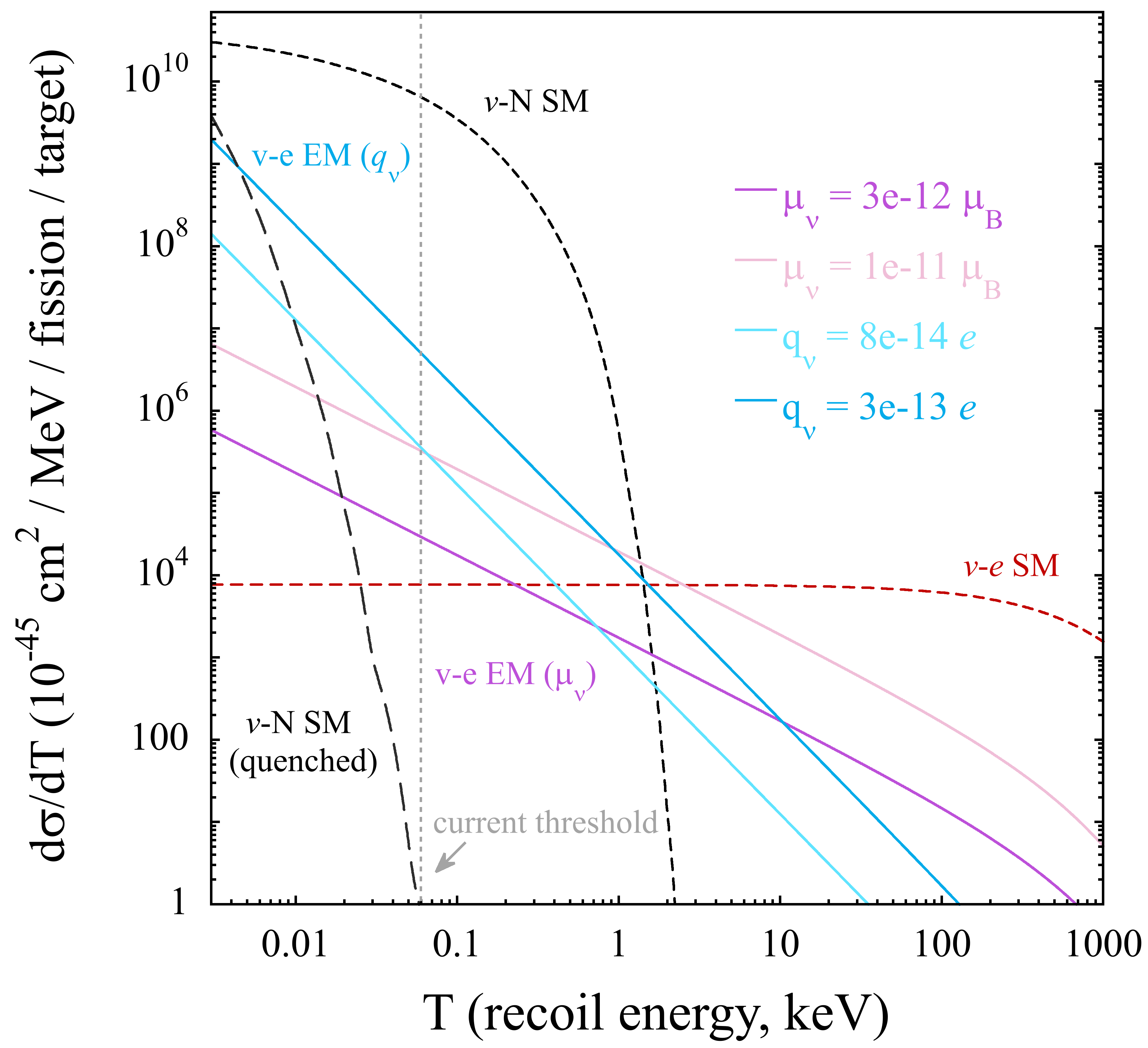}
    \caption{ Standard model (SM) and electromagnetic (EM) differential cross-sections for several values of neutrino magnetic moment $\mu_\nu$, in units of Bohr magneton $\mu_B$, and millicharge $q_\nu$, in units of elementary charge $e$, demonstrating that sensitivity to these channels is dominated by the lowest reachable recoil energies. Nuclear recoils are quenched according to \cite{csiqf_lowE}, with an uncertainty subdominant on the logarithmic scale of this figure. }
    \label{fig:diffxsec}
\end{figure}

This apparent weakness raises an opportunistic asymmetry: nuclear recoils are increasingly suppressed near threshold, while electron recoils are not.
Typical \cevns\ experiments at reactor sites \cite{miner,nucleus,richochet,conus,cogent,dresden2}, making use of the largest terrestrial flux of low-energy electron antineutrinos available, rely on the excess of nuclear recoils that spill over an energy threshold. This sensitivity to neutral-current $\bar{\nu}_{e} -N$ scattering events, and the inherent background difficulties of operating next to a reactor core, limit their sensitivity to physics beyond the standard model (BSM) that would similarly present as low-energy excesses \cite{dresden_em1,dresden_em2,reactor_em_3}.
Cryogenic CsI, which does not suffer from the same mixing of signals, is instead naturally optimized as a \ve\ detector in this regime.
Crucially, the competing backgrounds can also be minimized with local overburden and shielding.
The ideal deployment would be to the tendon gallery of a commercial reactor, where the reactor-correlated backgrounds become subdominant to the sub-ckkd (counts per keV-kg-day) reactor-independent backgrounds induced by environmental sources with modest overburden present \cite{cogent,phil}.

In this work, it is argued that a cryogenic CsI detector solves a specific problem in reactor neutrino physics: how to isolate a near-threshold \ve\ signal while suppressing the competing nuclear recoil channel.
Although limiting the \cevns\ program at low neutrino energies ($\mathcal{O}$(MeV)), this same property opens a powerful avenue for competitive background-limited searches in the field of exotic electromagnetic couplings of the neutrino.

    

    

\section{Neutrino-electron scattering}

Neutrino-electron scattering has long been the cleanest channel for probing BSM physics in the neutrino sector, as it is purely leptonic and has no hadronic uncertainties. Several experiments, notably TEXONO \cite{texono} and GEMMA \cite{gemma}, have set lasting limits using reactor antineutrinos as probes for deviations in the \ve\ scattering cross section. Although the SM cross section is small, it is precisely predicted, making any excesses at low electron recoil energies a clean signature of BSM effects. In this sub-keV region, since the measured low-energy QF strongly suppresses the visible \cevns\ contribution from reactor neutrinos (Fig. \ref{fig:diffxsec}), \ve\ becomes the dominant channel for CsI.

A measurement of any BSM effects is based on their contributions to the SM cross section. The differential cross section, following \cite{ve_orig}, can be written as:

\begin{equation} \label{eq:SMxsec}
\begin{split}
    \frac{d\sigma_{SM}}{dT} = \frac{2 G_F^2\, m_e c^2}{\pi}\,(\hbar c)^2\,
\left[g_R^{2} + g_L^{2}\left(1 - \frac{T}{E_\nu}\right)^{2} \right. \\ \left.
- \,\, g_L g_R \frac{m_e c^2 \, T}{E_\nu^{2}}
\right]
\end{split}
\end{equation}
where $g_L = \tfrac{1}{2} + \sin^2\theta_W$, $g_R = \sin^2\theta_W$, $E_\nu$ is the incoming neutrino energy, and $T$ the recoil energy. Among the four types of electromagnetic interactions, the additional contributions from the presence of a neutrino magnetic moment or electric charge can be sufficiently described, for present purposes, via additive terms:

\begin{equation}
    \frac{d\sigma_{EM}^{\mu_\nu}}{dT} = \frac{\pi\alpha^2}{m_e^2c^4} \, (\hbar c)^2 \, \left( \frac{\mu_\nu}{\mu_B} \right)^2 \left( \frac{1}{T} - \frac{1}{E_\nu} \right)
\end{equation}

\begin{equation}
    \frac{d\sigma_{EM}^{q_\nu}}{dT} = \frac{2\pi\alpha^2}{m_e^2c^4} \, (\hbar c)^2 \, \left( \frac{q_{\nu_e}^2}{T^2} \right)
\end{equation}
where the magnetic moment $\mu_\nu$ is expressed in units of Bohr magnetons ($\mu_B$), the neutrino millicharge $q_{\nu_e}$ in units of elementary charge $e$, and $\alpha$ is the fine structure constant. The other two types of interactions, the neutrino charge radius $\langle r_{\nu_e}^2 \rangle$ and anapole moment $a_{\nu_e}$, typically expressed in units of cm$^2$, can be described through the modification of the SM couplings in equation \ref{eq:SMxsec}, $g_L \rightarrow g_L^{EM}$ and $g_R \rightarrow g_R^{EM}$, via:

\begin{equation}
    g_{L/R}^{EM} = g_{L/R} \, + \, \frac{\sqrt{2}\pi\alpha}{(\hbar c)^2 G_F} \left( \frac{a_{\nu_e}}{18} + \frac{1}{3}\langle r_{\nu_e}^2 \rangle \right) \quad .
\end{equation}
The subsequent $\mathcal{O}(a_{\nu_e}^2)$ and $\mathcal{O}(\langle r_{\nu_e}^2 \rangle^2)$ terms of the modified total differential cross section are dropped for a compact and linearized form. The lack of inverse or inverse square dependence on electronic recoils greatly mitigates the impact of the threshold advantages offered by a cryogenic CsI detector. Parity in $a_{\nu_e}$ and $\langle r_{\nu_e}^2 \rangle$ sensitivity with higher-mass experiments can be reached as the separate advantage of radiopurity, discussed in section \ref{bckgr}, can justify smaller-scale setups, but not enough to be worth further comparison in this work.

\section{Conceptual design} \label{design}

\begin{figure}[t]
    \centering
    \includegraphics[width=1\linewidth]{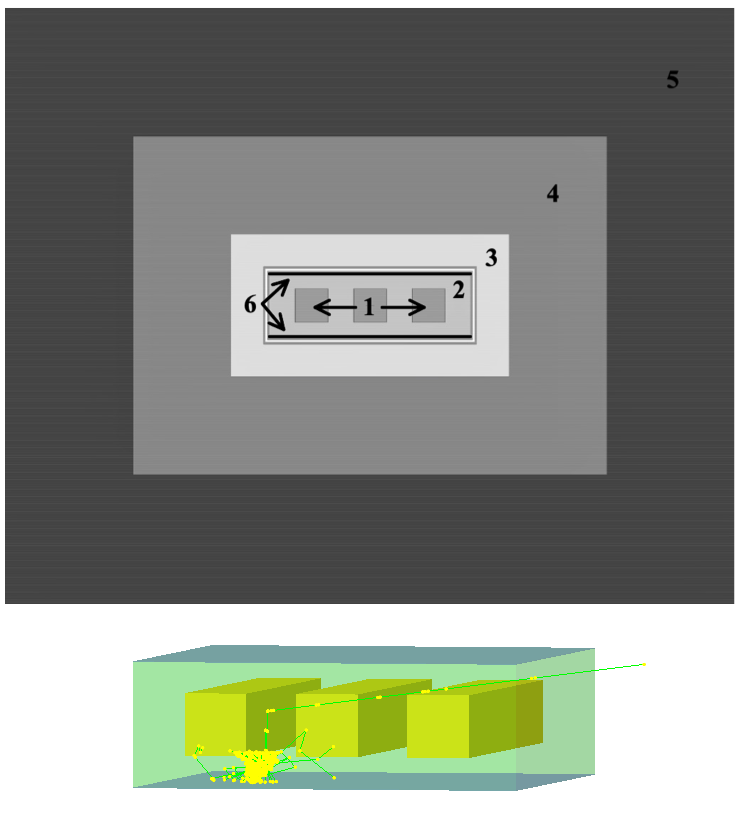}
    \caption{ {\it Top:} Cross section of the basic detector design and shield (see text). {\it Bottom:} Simulated internal background event escaping the crystal and interacting in the LAr volume. Light produced is then captured by the SiPM panels (see text). }
    \label{fig:assembly}
\end{figure}

The basic design of this experiment is displayed in Fig. \ref{fig:assembly} as a cross-section of the detector and shielding: 1) $\sim$10 kg of $5\times5\times32$ cm CsI crystals, 2) xenon-doped liquid argon (XeDLAr) reservoir and active veto, 3) inner plastic scintillator veto, 4) minimum 15 cm of Pb, 5) 20 cm of polyethylene, 6) two planes of silicon photosensors (in present calculations assumed to be silicon photomultipliers, or SiPMs), for monitoring of the LAr reservoir, covering a total area of $\sim$1000 cm$^2$. SiPMs like those described in \cite{sipm_darkside_rogers,sipm_darkside} have a sufficiently low dark count rate in cryogenic conditions to allow the operation of large-area tiles at single photoelectron sensitivity, while causing only a \%-scale dead time. A further 5 cm-thick outer muon veto of plastic scintillator (for five-sided coverage) is assumed for the work of the following sections. A high-efficiency, low-noise large area avalanche photodiode (LAAPD, \cite{bigapd}) at each end reads the scintillation light produced by radiation interactions in each of the CsI scintillators. Detecting mass could be increased, apart from the addition of crystal units, by increasing the cross-sectional area of scintillating crystals beyond the conservative dimensions used here and minimally adjusting the dimensions of the inner active veto volume \cite{kims_crys_size}. The total volume of the setup, including the shielding, is $\mathcal{O}(1)$ m$^3$.

The XeDLAr volume, besides being the thermal reservoir keeping the CsI crystals in the high-light yield regime, doubles as an active internal veto for the intrinsic backgrounds of the innermost materials in the setup (see section \ref{bckgr}). It, alongside the secondary plastic scintillator veto, neither of which needs to discriminate between nuclear and electron recoils, also serves to reinforce the efficiency of the external muon veto in tagging cosmic ray-induced events. The xenon-doping of liquid argon (LAr), above minimal concentrations of $\mathcal{O}(100)$ ppm, has been shown to act as an efficient wavelength shifter and to greatly increase the absolute light yield \cite{xedlar1}. The more abundant 178 nm scintillation photons also better match both the reflectance of internal materials, like a PTFE layer encasing the crystals, and the absorption regions of photosensors covering the veto volume. Any added complexity required to balance a cryogenic mixture is further offset by the increased edge-finding capabilities afforded by the reduced scintillation decay time.

Hygroscopic crystals such as CsI typically benefit from encapsulation for added ruggedness and isolation from trace amounts of water. Thin film fluoroplastics \cite{cytop1,teflon-af} transparent in the near-UV offer these advantages while remaining cryogenically stable. They can help simplify the internal geometry, including the suspension of the crystals and the reflector wrapping, while ensuring that any trace impurities in the noble liquid are chemically separated from the inorganic scintillator.


This detector combines several distinct, yet synergistic, developments over recent years aimed at exploiting the physics available in the low-energy regime. The high light yield, peaked at $\sim$340 nm, of undoped CsI at cryogenic temperatures has been noted by many groups \cite{amsler,mos1,mos2,nadeau,clark,liu,woody,zhang,mik}, along with its similar economic virtues as CsI[Na] \cite{NIMcenns}. Efforts in \cite{ESS,csiqf_lowE,lewis} confirmed that quantum efficiencies (QE) $\gtrsim80\%$ can be reached at these temperatures by combining novel wavelength shifters \cite{nol,nol2,nol3,nol4,nol5} with LAAPDs \cite{belle,jin} for few-photon thresholds. In parallel, LAAPDs with surface areas of 45 cm$^2$ have been operated with similar thresholds at 77 K \cite{bigapd}. Detectors made with low-resistivity (4 $\Omega \cdot$cm) silicon wafers were shown to boost the intrinsic QE of APDs to a similar $\sim$80\% in the near-UV \cite{highQEapd}. An ongoing partnership to reproduce and expand the capabilities of LAAPDs at short wavelengths is in progress with Fagor Electrónica \cite{fagor}. Both demonstrated avenues for increasing the effective QE of silicon detectors coupled to CsI crystals have the capability to lower the energy threshold enough to achieve the performance described in section \ref{sensitivity}. Each could also supplement or replace the SiPMs monitoring yet shorter wavelength scintillation light in the innermost active veto for more efficient background rejection.

\section{Backgrounds} \label{bckgr}

\begin{figure}[b]
    \centering
    \includegraphics[width=1\linewidth]{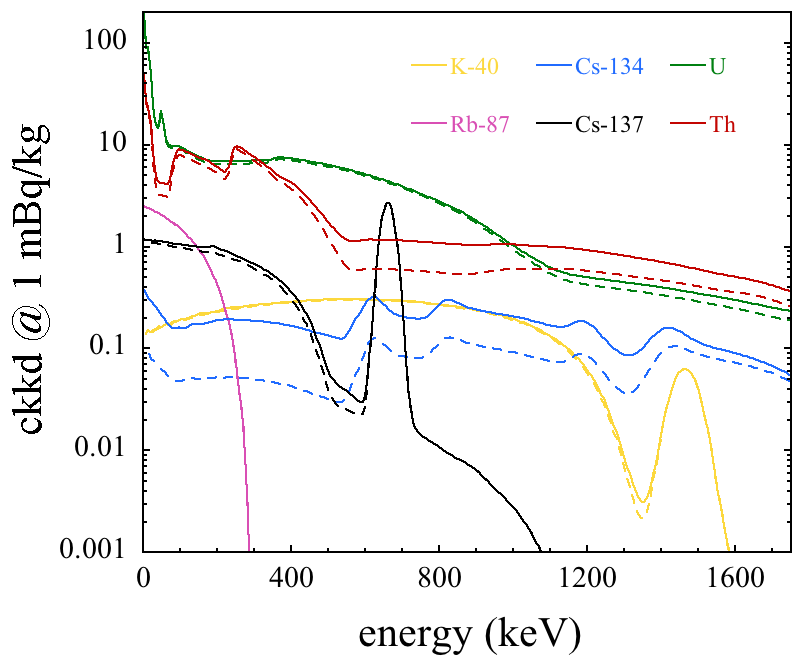}
    \caption{ Intrinsic backgrounds, obtained using GEANT4 simulations, in the CsI due to impurities. Assumes a fixed concentration of 1 mBq/kg per isotope, to provide a directly scalable response per contaminant. Solid lines denote total induced backgrounds and dashed lines illustrate the impact of the active inner veto. }
    \label{fig:activeveto}
\end{figure}

Operation in the tendon gallery of a commercial reactor offers a much more benign environment than activities within secondary containment \cite{phil,dresden1}. At the same time, one loses the timing benefits, {\it i.e.} anticoincident background subtraction, inherent to a spallation source. However, the steady-state environmental backgrounds can still be mitigated, with sufficient overburden and passive/active shielding, as demonstrated by previous reactor neutrino experiments \cite{cogent,dresden1,dresden2,phil,lewis}. The limiting contribution to the total background then becomes the intrinsic background of the internal components.

A subset of that background, the intrinsic $^{39}$Ar activity that would be present in the majority-LAr veto, is obviated by the mm-scale range of electrons in liquid and the relatively small mass ($\mathcal{O}(kg)$) within that range in direct contact with the reflector surrounding the crystals \cite{warp,DEAP_Ar}. Within the active veto, the bulk of these beta events, and most boundary ones, are taggable. Furthermore, at $\sim1$ Bq/kg activity, coincidence windows on the scale of $\mu$s contribute a negligible sub-\% deadtime.

Production of low-background CsI has been studied extensively by the KIMS collaboration \cite{kims_csi_wimp}. Techniques such as repeated dissolution in ``ultra-pure" water \cite{kims_bckgr} in powder extraction show that chemically similar contaminants to Cs are largely removable (limited only by water purity). Combined with purification via recrystallization \cite{recrystal}, drawing out Rb with each pass, a total internal background of $\sim$2.5 ckkd in Tl-doped CsI was reached even with suboptimal purification and coupling to photomultiplier tubes (PMTs) \cite{kims_csi_2025}.

An exception to the purification process, save in the use of salts previously stored underground, is the production of $^{134}$Cs by neutron capture in the stable isotope. As demonstrated in Fig. \ref{fig:activeveto}, for a given mock-activity, a thermal bath that doubles as an active veto is an efficient means of rejection for such isotopes with decays featuring correlated gammas escaping the crystal (bottom panel of Fig. \ref{fig:assembly}). GEANT4 version 11 \cite{g4} was used for optical simulations to help quantify this efficiency by mapping the position-dependent transport of scintillation light to the two submerged large-area SiPM arrays \cite{xedlar2,idx_refrac}. Simulations of the full response of the system to internal radioactivity assume a conservative light yield of 60 photons/keV for XeDLAr and 75\% majority-diffuse reflectance of PTFE at this wavelength \cite{xedlar1,ptfe_xe_vacuum,ptfe_xe_liquid}. Photosensors monitoring that volume are assumed to have a QE $= 30\%$ at xenon emission wavelengths (that is, excluding the added complexity of waveshifters or a custom sensor production). For an experiment in the low-energy regime, this background can be approximated as linear over a small (keV-scale) energy region.

\begin{figure}
    \centering
    \includegraphics[width=1\linewidth]{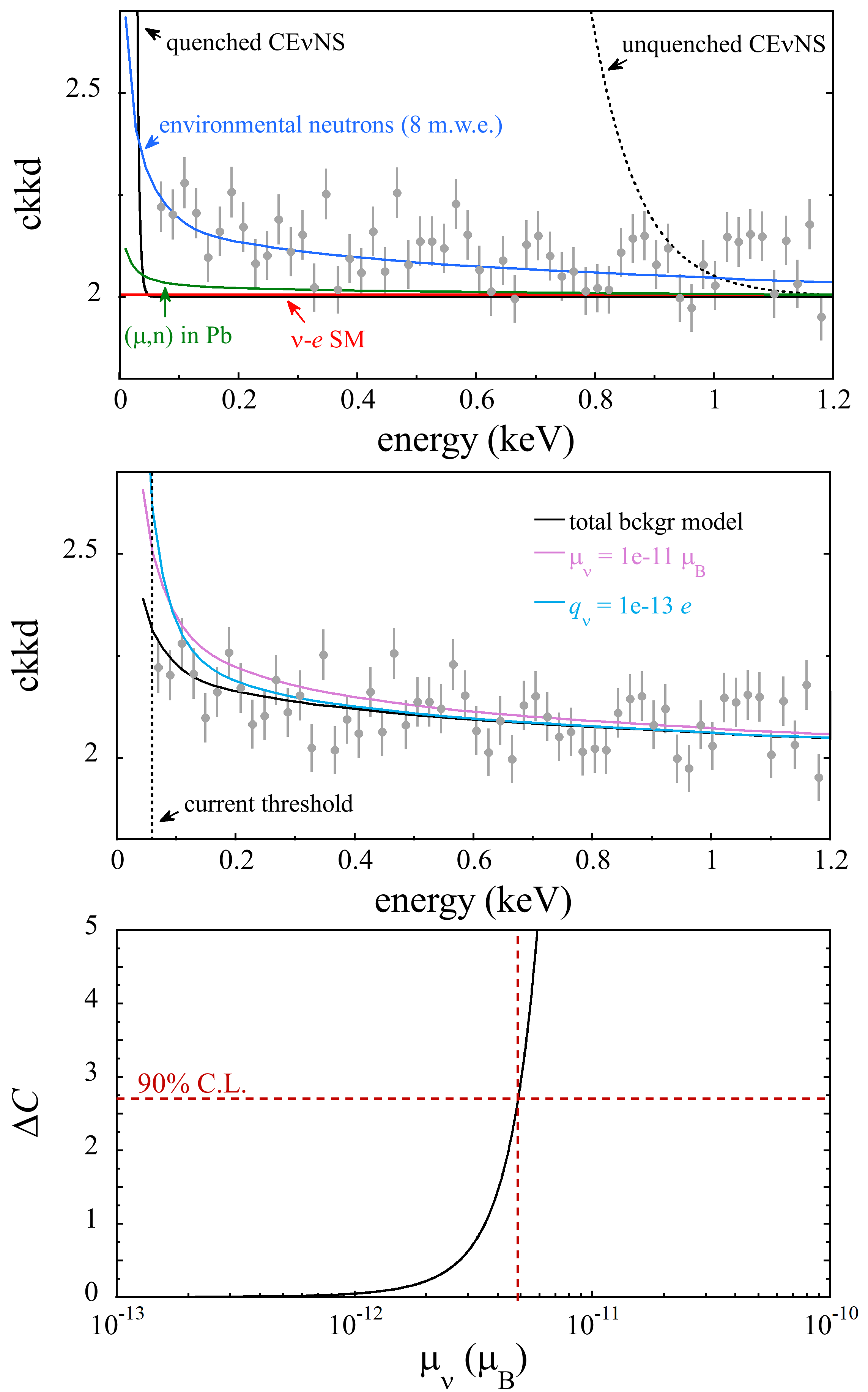}
    \caption{ {\it Top:} Representative simulated example of a dataset in counts/keV/kg/day consisting of 40 kg of CsI exposed to a neutrino flux of $2.7\times10^{13}$ $\bar{\nu_e}$/cm$^2$/s (the flux experienced by GEMMA \cite{gemma}) over 2 years of data taking. Contributing electron and nuclear recoil spectra (quenched) at 8 m.w.e. overburden are also shown. {\it Mid:} Same Monte Carlo dataset overlaid with example one-dimensional EM contributions from a neutrino magnetic moment $\mu_\nu$ and electronic charge $q_\nu$, a lack of which is distinguishable against the null (black contour) above a 60 eV threshold. {\it Bot:} One-dimensional distribution of the likelihood $\Delta C$ for the representative dataset with the $90\%$ C.L. boundary of $\mu_\nu$. }
    \label{fig:mc}
\end{figure}

The simulated response of the detector to environmental neutrons (cosmic-ray tertiaries and those from fission decays and ($\alpha,n$) reactions) and the tertiary neutrons produced by muons (cosmic-ray secondaries) traversing the detector and shielding geometries was found using the MCNP-Polimi code \cite{polimi}. The hardness spectra adopted for environmental neutrons are described in \cite{env_neutrons_1,env_neutrons_2}- below 20 MeV they were simulated assuming an isotropic origin and above that energy with an originating skyward bias \cite{env_n_angles}. Neutrons from $(\mu,n)$ reactions \cite{muon_neutrons_1,muon_neutrons_2}) were distributed homogeneously in the lead shielding. Their contribution to the background spectrum is mainly reduced by the overall tagging efficiency presumed of the external muon veto ($99.75\%$), which is comparable to the performance of a similar veto used for the NCC-1701 detector (or Dresden-II) \cite{dresden1}. The various components of the total background model are demonstrated in the top panel of Fig. \ref{fig:mc} overlaid on an example intrinsic background that contributes 2 ckkd. It is evident that with the current configuration, even modest overburden renders environmental sources subdominant. This establishes a realistic background floor low enough to expose near-threshold \ve\ signals for the sensitivity estimates presented in the following section.


\section{Sensitivity projections} \label{sensitivity}

\begin{figure*}
\centering
\includegraphics[width=\textwidth]{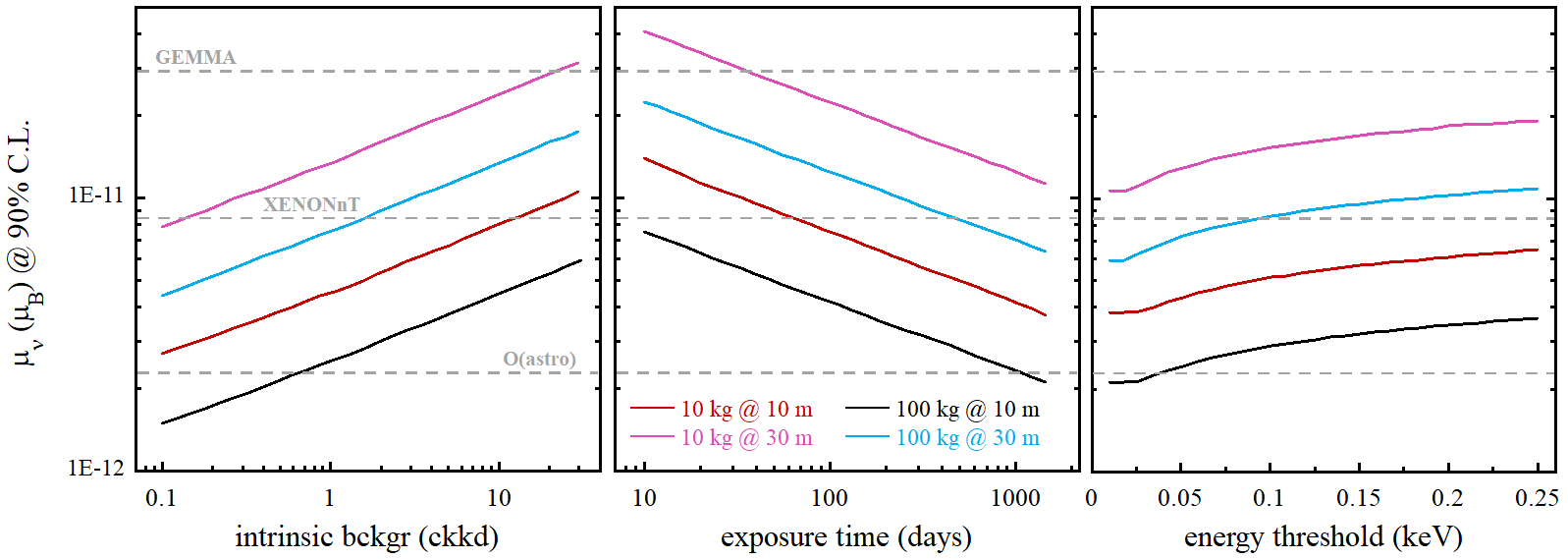}
\caption{\label{fig:mu} { Simulated 90\% C.L. limits on the electron-neutrino magnetic moment projected for a sampling of detector parameters: magnitude of the dominant intrinsic background, energy threshold, and exposure time to a given neutrino flux. See text for details. }  }
\end{figure*}

Having established that cryogenic CsI suppresses the nuclear-recoil channel while maintaining sensitivity to near-threshold electron recoils, we now quantify the resulting sensitivity to neutrino electromagnetic interactions.
A total background model was built from the contributions of the components described above. Due to the limited understanding of the resolution at these sub-keV energies and the low scintillation photon statistics involved in the near-threshold region of interest ($<$10 PE), the individual components of the model were each smeared with a Poisson. All sources of nuclear recoils in the crystals were quenched according to the findings of \cite{csiqf_lowE}. Each trial of the Monte Carlo, after detector parameter selection, simulated a full dataset with a null signal by sampling this contour. Fig. \ref{fig:mc} describes one of these trials for a detector with a CsI mass of 40 kg, an energy threshold of 60 eV ($\sim$6 PE), and an intrinsic background of 2 ckkd placed 13.9 m from a 3 GW$_{th}$ reactor core with 8 m.w.e. overburden. This mimics the running conditions of GEMMA \cite{gemma} in the keV-scale energy region, save an artificial lightening of the 70 m.w.e. overburden there to better visualize the impact of environmental and cosmogenic-induced neutrons.
The inverse dependence on recoil energy of neutrino electromagnetic interactions would have them present as statistically significant excesses close to threshold, visible when contrasted against the expectation via a likelihood function.
As a sanity check, the Monte Carlo analysis described here was also performed with a 1.5 kg Ge detector above a 2.8 keV energy threshold, following the spectra given in \cite{gemma}, to recover the same $\mu_{\nu_e}$ sensitivity presented there.

The typical methodology isolates the neutrino-induced component by comparing the spectra taken during the reactor operation (ON) and refueling (OFF) periods for the residual signal.
Beyond that, statistical uncertainties will drive the sensitivity above a given threshold. ON and OFF spectra can be compared bin-by-bin \cite{texono_result_1,gemma_first}, but to combat the dominating uncertainty of the shorter OFF period of a reactor, additional information can be introduced. Using a smooth function to parametrize the region of interest (ROI) of the OFF spectra before comparison renders this subdominant over the statistical uncertainty in the ON periods \cite{gemma_3yr}. The theoretical model of the background fills the role of an expectation function for each trial in this work.

To estimate each of the new physics parameters and conservatively avoid binning-dependent trends across a wide range of bin statistics, exposure times, and detector parameters, the Cash statistic $C$ was adopted from the Poisson likelihood:

\begin{equation}
    C = 2\sum^N_{i=1} ( M_i(x) - D_i\ln{M_i(x)} )
\end{equation}

where, for each bin $i$, $D$ defines the Monte Carlo data and $M(x)$ defines the model with extra electromagnetic parameter $x$ (be it $\mu_{\nu_e}$ or $q_{\nu_e}$). 
This is visualized in the bottom panel of Fig. \ref{fig:mc} for the representative trial dataset. By the profile-likelihood rule for one parameter $\Delta C \approx \Delta \chi ^2$ and the 90\% confidence level (CL) is at $\Delta C = 2.71$ \cite{wilks}. As the expressions for these additional physics parameters were linearized in each parameter, $\Delta C$ is a distribution of the absolute value only. Taking the average of many trials of the Monte Carlo and analysis for the same set of detector conditions gives an estimate of the achievable sensitivity with existing detector capabilities.

\begin{figure*}[t]
\centering
\includegraphics[width=\textwidth]{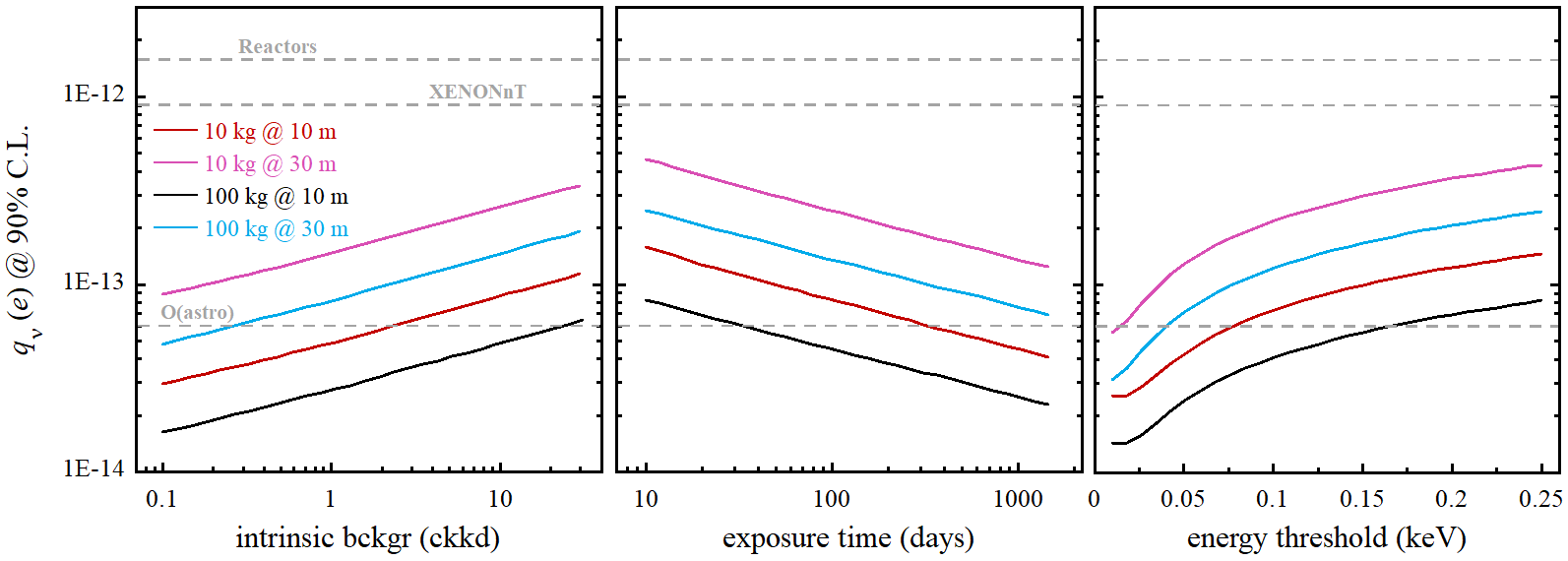}
\caption{\label{fig:q} { Simulated 90\% C.L. limits on the electron-neutrino electric charge projected for a sampling of detector parameters: magnitude of the dominant intrinsic background, energy threshold, and exposure time to a neutrino flux. See text for details. }  }
\end{figure*}

Fig. \ref{fig:mu} shows the extracted 90\% C.L. limits of these analyses for the neutrino magnetic moment when all other new physics parameters are set to zero. Increasing distance from the core is not completely nullified by increasing mass, as the backgrounds are independent of the neutrino flux. Fig. \ref{fig:q} repeats this exercise for the neutrino millicharge. Unless otherwise specified, contours reference a 2-year exposure with sufficient knowledge of the shape of the background. Likewise, unless variable, a gallery-typical overburden of 20 m.w.e., an intrinsic internal background of 1 ckkd, an energy threshold of 60 eV, and a reactor power of 3 GW$_{th}$ are assumed. As Fig. \ref{fig:mu} demonstrates, an order of magnitude improvement over the current best limits set with reactor neutrinos (GEMMA's $\mu_{\nu_e} < 2.9 \times 10^{-11} \mu_B$) and a clear improvement on the best-overall limit for this flavor set by XENONnT ($\mu_{\nu_e} < 8.5 \times 10^{-12} \mu_B$) is achievable with this technique even with low sensitive masses. The neutrino electric charge (Fig. \ref{fig:q}), with its stronger inverse-square dependence on recoil energy, sees a sensitivity boost of 1-2 orders of magnitude beyond the current state of the art from reactor experiments and XENONnT \cite{Khan_2023}. These sensitivities place this technique within the regime of limits set by model-dependent astrophysical observations \cite{Raffelt_1999,Davidson_2000} and open the door for macro-scale constraints from low-mass experiments.

\section{Summary and conclusions}

It has been shown that cryogenic CsI occupies a distinctive niche in reactor neutrino physics.
By combining an intrinsic suppression of nuclear recoils with a high scintillation yield at cryogenic temperatures, pure CsI is advantageous for background-limited, near-threshold searches in the electronic recoil channel.
Combined with the low and stable background environment of reactor tendon galleries, this makes the detector concept tailor-made for \ve\ physics rather than a generic low-threshold technology.
With existing technologies and detector performance, order-of-magnitude improvements to the present limits from reactor antineutrinos experiments \cite{gemma,texono_result_1,Khan_2023} are achievable with target masses of only tens of kilograms.
This reaches parameter space normally only associated with much larger experiments and
enters the $\mathcal{O}(10^{-14}-10^{-13})$ and $\mathcal{O}(10^{-12})$ regions probed by astrophysical constraints for $q_{\nu_e}$ and $\mu_{\nu_e}$, respectively \cite{Aprile_2022,Agostini_2017,Giunti_2015}.
A dedicated effort to further the ability to grow pure scintillator stock, free of radiopurities and without the constraint of large-scale commercialization, could support intrinsic backgrounds better than $\mathcal{O}(1)$ ckkd \cite{kims_bckgr} and a corresponding improvement in the viability of this detector technology.


Electromagnetic properties of neutrinos are a prediction of many models incorporating finite neutrino masses, and improving the experimental sensitivity to magnetic moment and millicharge interactions directly tests this physics at the laboratory scale. As demonstrated in this work, new physics is reachable even with kg-scale detectors over mere week-scale exposure times. A cryogenic CsI experiment in a reactor tendon gallery thus offers a uniquely promising and scalable path toward probing BSM physics in the present day.

\begin{acknowledgement}
The author would like to thank S. Torelli, S.\,R. Soleti, F. Monrabal, and J.\,J. Gomez-Cadenas for feedback on earlier drafts of this article.
\end{acknowledgement}
%

%
\bibliographystyle{unsrt_notitle} 
\bibliography{apssamp}

@PREAMBLE{
 "\providecommand{\noopsort}[1]{}" 
 # "\providecommand{\singleletter}[1]{#1}%" 
}

@article{Davidson_2000,
   title={Updated bounds on milli-charged particles},
   volume={2000},
   ISSN={1029-8479},
   url={http://dx.doi.org/10.1088/1126-6708/2000/05/003},
   DOI={10.1088/1126-6708/2000/05/003},
   number={05},
   journal={JHEP},
   publisher={Springer Science and Business Media LLC},
   author={Davidson, Sacha and Hannestad, Steen and Raffelt, Georg},
   year={2000},
   month=may, pages={003–003} }

@article{Raffelt_1999, title={Limits on neutrino electromagnetic properties — an update}, volume={320}, ISSN={0370-1573}, url={https://www.sciencedirect.com/science/article/pii/S0370157399000745}, DOI={https://doi.org/10.1016/S0370-1573(99)00074-5}, abstractNote={Limits on neutrino electromagnetic properties from laboratory experiments and astrophysical arguments are reviewed with an emphasis on the currently favored range of small neutrino masses. We derive a helioseismological limit on the charge and dipole moment for all flavors of eν≲6×10−14e and μν≲4×10−10μB (Bohr magneton). The most restrictive limits remain those from the plasmon decay in globular-cluster stars of eν≲2×10−14e and μν≲3×10−12μB.}, number={1}, journal={Phys. Rep.}, author={Raffelt, Georg G.}, year={1999}, pages={319–327} }

@article{jparc_pheno,
      title={Coherent Elastic Neutrino-Nucleus Scattering at the Japan Proton Accelerator Research Complex}, 
      author={J. I. Collar and others},
      year={2026},
      volume={2026},
      number={5},
      month={May},
      pages={106},
      eprint={2512.19788},
      archivePrefix={arXiv},
      primaryClass={hep-ph},
      url={https://doi.org/10.1007/JHEP05(2026)106}, 
      journal={JHEP},
}

@article{csiqf_lowE,
      title={Scintillation response of cryogenic CsI to few-keV and sub-keV nuclear recoils}, 
      author={J. I. Collar and C. M. Lewis and A. Simón and S. G. Yoon},
      eprint={2512.09820},
      note ={2512.09820},
      archivePrefix={arXiv}, 
  journal = {Phys. Rev. C},
  volume = {114},
  issue = {3},
  pages = {034607},
  numpages = {7},
  year = {2026},
  month = {Sep},
  publisher = {APS},
  doi = {10.1103/gkdb-l6v8},
  url = {https://link.aps.org/doi/10.1103/gkdb-l6v8}
}

@article{xedlar2,
   title={Large-scale, precision xenon doping of liquid argon},
   volume={1011},
   ISSN={0168-9002},
   url={http://dx.doi.org/10.1016/j.nima.2021.165575},
   DOI={10.1016/j.nima.2021.165575},
   journal={Nucl. Instr. Meth. A},
   publisher={Elsevier BV},
   author={McFadden, N. and others},
   year={2021},
   month={Sept},
   pages={165575} }

@misc{idx_refrac,
      title={Index of refraction, Rayleigh scattering length, and Sellmeier coefficients in solid and liquid argon and xenon}, 
      author={Emily Grace and Alistair Butcher and Jocelyn Monroe and James A. Nikkel},
      year={2017},
      eprint={1502.04213},
      archivePrefix={arXiv},
      note={arXiv:1502.04213},
      primaryClass={physics.ins-det},
      url={https://arxiv.org/abs/1502.04213}, 
}

@article{ptfe_xe_vacuum,
   title={Reflectance of polytetrafluoroethylene for xenon scintillation light},
   volume={107},
   ISSN={1089-7550},
   url={http://dx.doi.org/10.1063/1.3318681},
   DOI={10.1063/1.3318681},
   number={6},
   journal={J. Appl. Phys.},
   publisher={AIP Publishing},
   author={Silva, C. and Pinto da Cunha, J. and Pereira, A. and Chepel, V. and Lopes, M. I. and Solovov, V. and Neves, F.},
   year={2010},
   month={Mar},
   }

@article{ptfe_xe_liquid,
doi = {10.1088/1748-0221/12/01/P01017},
url = {https://doi.org/10.1088/1748-0221/12/01/P01017},
year = {2017},
month = {Jan},
volume = {12},
number = {01},
pages = {P01017},
author = {Neves, F. and others},
title = {Measurement of the absolute reflectance of polytetrafluoroethylene (PTFE)  immersed in liquid xenon},
journal = {JINST}
}

@article{warp, title={Measurement of the specific activity of 39Ar in natural argon}, volume={574}, ISSN={0168-9002}, DOI={https://doi.org/10.1016/j.nima.2007.01.106}, abstractNote={We report on the measurement of the specific activity of 39Ar in natural argon. The measurement was performed with a 2.3-l two-phase (liquid and gas) argon drift chamber. The detector was developed by the WARP Collaboration as a prototype detector for WIMP Dark Matter searches with argon as a target. The detector was operated for more than two years at Laboratori Nazionali del Gran Sasso, Italy, at a depth of 3400m w.e. The specific activity measured for 39Ar is 1.01±0.02(stat)±0.08(syst)Bq per kg of natAr.}, number={1}, journal="Nucl. Instr. Meth. A", author={Benetti, P. and others}, year={2007}, pages={83–88} }

@article{DEAP_Ar, title={Precision measurement of the specific activity of $$^{39}$$Ar in atmospheric argon with the DEAP-3600 detector}, volume={83}, ISSN={1434-6052}, DOI={10.1140/epjc/s10052-023-11678-6}, abstractNote={The specific activity of the $$beta $$decay of $$^{39}$$Ar in atmospheric argon is measured using the DEAP-3600 detector. DEAP-3600, located 2 km underground at SNOLAB, uses a total of (3269 ± 24) kg of liquid argon distilled from the atmosphere to search for dark matter. This detector is well-suited to measure the decay of $$^{39}$$Ar owing to its very low background levels. This is achieved in two ways: it uses low background construction materials; and it uses pulse-shape discrimination to differentiate between nuclear recoils and electron recoils. With 167 live-days of data, the measured specific activity at the time of atmospheric extraction is (0.964 ± 0.001$$_textrm{stat}$$± 0.024$$_textrm{sys}$$) Bq/kg$$_textrm{atmAr}$$, which is consistent with results from other experiments. A cross-check analysis using different event selection criteria and a different statistical method confirms the result.}, number={7}, journal={Eur. Phys. J. C}, author={Adhikari, P. and DEAP Collaboration}, year={2023}, month={july}, pages={642}, language={en} }

@article{dresden1,
  title = {First results from a search for coherent elastic neutrino-nucleus scattering at a reactor site},
  author = {Colaresi, J. and Collar, J. I. and Hossbach, T. W. and Kavner, A. R. L. and Lewis, C. M. and Robinson, A. E. and Yocum, K. M.},
  journal = {Phys. Rev. D},
  volume = {104},
  issue = {7},
  pages = {072003},
  numpages = {8},
  year = {2021},
  month = {Oct},
  publisher = {American Physical Society},
  doi = {10.1103/PhysRevD.104.072003},
  url = {https://link.aps.org/doi/10.1103/PhysRevD.104.072003}
}

@article{dresden2,
  title = {Measurement of Coherent Elastic Neutrino-Nucleus Scattering from Reactor Antineutrinos},
  author = {Colaresi, J. and Collar, J. I. and Hossbach, T. W. and Lewis, C. M. and Yocum, K. M.},
  journal = {Phys. Rev. Lett.},
  volume = {129},
  issue = {21},
  pages = {211802},
  numpages = {6},
  year = {2022},
  month = {Nov},
  publisher = {American Physical Society},
  doi = {10.1103/PhysRevLett.129.211802},
  url = {https://link.aps.org/doi/10.1103/PhysRevLett.129.211802}
}

@article{science,
	author = {Akimov, D. and others},
	volume = {357},
	number = {6356},
	pages = {1123--1126},
	year = {2017},
	doi = {10.1126/science.aao0990},
	publisher = {American Association for the Advancement of Science},
	issn = {0036-8075},
	journal = {Science},
    eprint = "1708.01294 ",
}

@book{bjorn,
 author = {Scholz, Bjorn},
 date = {2018},
 doi = {10.1007/978-3-319-99747-6},
 series = {Springer Theses},
 publisher = {Springer Cham},
 editor = {},
 note ={1904.01155},
 eprint={1904.01155},
 pages = {},
 title = {First Observation of Coherent Elastic Neutrino-Nucleus Scattering},
 url = {https://app.dimensions.ai/details/publication/pub.1108429504},
 year = {2018}
}

@article{NIMcenns,
journal = "Nucl. Instr. Meth. A",
volume = "773",
pages = "56 - 65",
year = "2015",
issn = "0168-9002",
doi = "https://doi.org/10.1016/j.nima.2014.11.037",
url = "http://www.sciencedirect.com/science/article/pii/S0168900214013254",
author = " Collar, J. I. and others",
  eprint = "1407.7524",
}

@phdthesis{nicole,
author={Fields, Nicole E.},
school={U. of Chicago},
year={2014},   
}

@phdthesis{prior,
author={Prior, Charles G.},
school={Duke University},
year={2025},   
}

@article{ESS,
    author = "Baxter, D. and others",
    eprint = "1911.00762",
    archivePrefix = "arXiv",
    reportNumber = "IFIC/19-45, YITP-SB-19-37, FERMILAB-PUB-19-612-V",
    doi = "10.1007/JHEP02(2020)123",
    journal = "JHEP",
    volume = "02",
    pages = "123",
    year = "2020",
}

@article{csiqf,
  author = {Lewis, C. M. and Collar, J. I.},
  journal = {Phys. Rev. C},
  volume = {104},
  issue = {1},
  pages = {014612},
  numpages = {7},
  year = {2021},
  month = {Jul},
  publisher = {American Physical Society},
  doi = {10.1103/PhysRevC.104.014612},
  url = {https://link.aps.org/doi/10.1103/PhysRevC.104.014612},
    eprint = "2101.03264",
}

@article{chireactor,
    author = "Wang, Lei and others",
    eprint = "2212.11515",
    archivePrefix = "arXiv",
    doi = "10.1140/epjc/s10052-024-12800-y",
    journal = "Eur. Phys. J. C",
    volume = "84",
    number = "4",
    pages = "440",
    year = "2024"
}

@article{cohcsi,
    author = "Barbeau, P. S. and others",
    eprint = "2311.13032",
    archivePrefix = "arXiv",
     doi = "10.1103/PhysRevD.109.092005",
    journal = "Phys. Rev. D",
    volume = "109",
    number = "9",
    pages = "092005",
    year = "2024"
}

@article{clovers,
    author = "Su, Chenguang and Liu, Qian and Liang, Tianjiao",
    eprint = "2303.13423",
    archivePrefix = "arXiv",
    doi = "10.3390/psf2023008019",
    journal = "Phys. Sci. Forum",
    volume = "8",
    number = "1",
    pages = "19",
    year = "2023"
}

@article{kims_csi_2025,
    author = "Kim, W. K. and others",
    title = "{Scintillation characteristics of an undoped CsI crystal at low-temperature for dark matter search}",
    eprint = "2312.07957",
    archivePrefix = "arXiv",
    primaryClass = "physics.ins-det",
    doi = "10.1016/j.astropartphys.2025.103150",
    journal = "Astropart. Phys.",
    volume = "173",
    pages = "103150",
    year = "2025"
}

@article{kims_bckgr,title={Development of low-background CsI(T) crystals for WIMP search}, volume={571}, rights={https://www.elsevier.com/tdm/userlicense/1.0/}, ISSN={0168-9002}, url={https://linkinghub.elsevier.com/retrieve/pii/S0168900206022650}, DOI={10.1016/j.nima.2006.10.325}, abstractNote={Search for weakly interacting massive particles (WIMPs) is being carried out at the underground laboratory, Yangyang, Korea. Characteristics and internal background of CsI(T‘) crystal have been investigated. In our extensive R&D, we reduced internal background in the CsIðT‘Þ crystal. With the latest, we have achieved 5:50 # 0:10 cpd (counts/keV/kg/day) at 10–15 keV low-energy region. Further reduction of internal background is foreseen with the CsI powder lately produced.}, number={3}, journal={Nucl. Instr. Meth. A}, publisher={Elsevier BV}, author={Lee, H.S. and others}, year={2007}, month=feb, pages={644–650} }

@article{kims_csi_wimp, title={Study of the internal background of CsI(Tℓ) crystal detectors for dark matter search}, volume={500}, ISSN={01689002}, DOI={10.1016/S0168-9002(03)00346-2}, abstractNote={A search for particle cold dark matter with CsIðTcÞ crystal is being prepared at the Cheong-Pyeong underground laboratory in Korea. The background spectra of CsIðTcÞ crystal detectors in a prototype shield were obtained. The lowest background count rate of the test sample of crystals is measured to be 64:775:1 counts/keV/kg/day in the energy range of 5–20 keV: Quantitative estimation of residual radioactive isotope in CsIðTcÞ was made using the GEANT4 Monte Carlo simulation. Analysis results show that CsIðTcÞ crystal could be a good candidate for direct detection of WIMPs when the contamination level of cesium radioisotopes is reduced to under a few mBq/kg.}, number={1–3}, journal={Nucl. Instr. Meth. A}, author={Kim, T.Y and others}, year={2003}, month=mar, pages={337–344} }

@article{kims_crys_size,
  title = {Search for low-mass dark matter with CsI(Tl) crystal detectors},
  author = {Lee, H. S. and others},
  collaboration = {KIMS Collaboration},
  journal = {Phys. Rev. D},
  volume = {90},
  issue = {5},
  pages = {052006},
  numpages = {6},
  year = {2014},
  month = {Sep},
  publisher = {American Physical Society},
  doi = {10.1103/PhysRevD.90.052006},
  url = {https://link.aps.org/doi/10.1103/PhysRevD.90.052006}
}

@article{recrystal,
title = {Effect of recrystallisation on the radioactive contamination of CaWO4 crystal scintillators},
journal = {Nucl. Instr. Meth. A},
volume = {631},
number = {1},
pages = {44-53},
year = {2011},
issn = {0168-9002},
doi = {https://doi.org/10.1016/j.nima.2010.11.118},
url = {https://www.sciencedirect.com/science/article/pii/S0168900210026653},
author = {F.A. Danevich and others}
}

@article{gemma_3yr, title={GEMMA experiment: three years of the search for the neutrino magnetic moment}, volume={7}, ISSN={1547-4771, 1531-8567}, DOI={10.1134/S1547477110060063}, abstractNote={The result of the 3-year neutrino magnetic moment measurement at the Kalinin Nuclear Power Plant with the GEMMA spectrometer is presented. Antineutrino-electron scattering is investigated. A high-purity germanium detector of 1.5 kg placed at a distance of 13.9 m from the 3 GW(th) reactor core is used in the spectrometer. The antineutrino flux is 2.7E13 1/scm/s. The differential method is used to extract (nu-e) electromagnetic scattering events. The scattered electron spectra taken in 5184+6798 and 1853+1021 hours for the reactor ON and OFF periods are compared. The upper limit for the neutrino magnetic moment < 3.2E-11 Bohr magneton at 90% CL is derived from the data processing.}, note={arXiv:0906.1926 [hep-ex]}, number={6}, journal={Phys. Par. Nucl. Lett.}, author={Beda, A. G. and others}, year={2010}, month=nov, pages={406–409} }

@article{gemma_first, title={The first result of the neutrino magnetic moment measurement in the GEMMA experiment}, volume={70}, ISSN={1063-7788, 1562-692X}, DOI={10.1134/S1063778807110063}, abstractNote={The first result of the neutrino magnetic moment measurement at the Kalininskaya Nuclear Power Plant (KNPP) with the GEMMA spectrometer is presented. An antineutrino-electron scattering is investigated. A high-purity germanium detector of 1.5 kg placed 13.9 m away from the 3 GW reactor core is used in the spectrometer. The antineutrino flux is $2.73times 10^{13} nu_e / cm^2 / s$. The differential method is used to extract the $nu$-e electromagnetic scattering events. The scattered electron spectra taken in 6200 and 2064 hours for the reactor ON and OFF periods are compared. The upper limit for the neutrino magnetic moment $mu_nu < 5.8times 10^{-11}$ Bohr magnetons at 90{%} CL is derived from the data processing.}, note={arXiv:0705.4576 [hep-ex]}, number={11}, journal={Phys. Atom. Nucl.}, author={Beda, A. G. and others}, year={2007}, month=nov, pages={1873–1884} }

@article{texono_result_1, title={Measurement of $\nu$-electron scattering cross section with a CsI(Tl) scintillating crystal array at the Kuo-Sheng nuclear power reactor}, volume={81}, ISSN={1550-7998, 1550-2368}, DOI={10.1103/PhysRevD.81.072001}, number={7}, journal={Phys. Rev. D}, author={Deniz, M. and others}, year={2010}, month=apr, pages={072001} }

@phdthesis{lewis,
    author = "Lewis, C. M.",
    title = "{Particle physics in the sub-keV energy regime}",
    note = "arXiv:2310.01314",
    archivePrefix = "arXiv",
    primaryClass = "physics.ins-det",
    doi = "10.6082/uchicago.5730",
    school = "U. of Chicago",
    year = "2023"
}

@article{nucleus,
   title={The $\nu$-cleus experiment: a gram-scale fiducial-volume cryogenic detector for the first detection of coherent neutrino–nucleus scattering},
   volume={77},
   ISSN={1434-6052},
   url={http://dx.doi.org/10.1140/epjc/s10052-017-5068-2},
   DOI={10.1140/epjc/s10052-017-5068-2},
   number={8},
   journal={Eur. Phys. J. C},
   publisher={Springer Science and Business Media LLC},
   author={Strauss, R. and others},
   year={2017},
   month=jul
}

@article{richochet,
doi = {10.1088/1361-6471/aa83d0},
url = {https://dx.doi.org/10.1088/1361-6471/aa83d0},
year = {2017},
month = {aug},
publisher = {IOP Publishing},
volume = {44},
number = {10},
pages = {105101},
author = {Billard, J and others},
title = {Coherent neutrino scattering with low temperature bolometers at Chooz reactor complex},
journal = {J. Phys. G}
}

@article{miner,
title = {Background studies for the MINER Coherent Neutrino Scattering reactor experiment},
journal = {Nucl. Instrum. Meth. A},
volume = {853},
pages = {53-60},
year = {2017},
issn = {0168-9002},
doi = {https://doi.org/10.1016/j.nima.2017.02.024},
url = {https://www.sciencedirect.com/science/article/pii/S0168900217302085},
author = {G. Agnolet and others}
}

@article{dresden_em1,
   title={Impact of the Dresden-II and COHERENT neutrino scattering data on neutrino electromagnetic properties and electroweak physics},
   volume={2022},
   ISSN={1029-8479},
   url={http://dx.doi.org/10.1007/JHEP09(2022)164},
   DOI={10.1007/jhep09(2022)164},
   number={9},
   journal={JHEP},
   publisher={Springer Science and Business Media LLC},
   author={Atzori Corona, M. and others},
   year={2022},
   month=sep }

@article{dresden_em2,
   title={Bounds on new physics with data of the Dresden-II reactor experiment and COHERENT},
   volume={2022},
   ISSN={1029-8479},
   url={http://dx.doi.org/10.1007/JHEP05(2022)037},
   DOI={10.1007/jhep05(2022)037},
   number={5},
   journal={JHEP},
   publisher={Springer Science and Business Media LLC},
   author={Coloma, Pilar and Esteban, Ivan and Gonzalez-Garcia, M. C. and Larizgoitia, Leire and Monrabal, Francesc and Palomares-Ruiz, Sergio},
   year={2022},
   month=may }

@article{texono,
  author = {Singh, L. and others},
  journal = {Phys. Rev. D},
  volume = {99},
  issue = {3},
  pages = {032009},
  numpages = {8},
  year = {2019},
  month = {Feb},
  publisher = {Amer. Phys. Soc.},
  doi = {10.1103/PhysRevD.99.032009},
  url = {https://link.aps.org/doi/10.1103/PhysRevD.99.032009}
}

@article{conus,
	doi = {10.1088/1742-6596/1342/1/012094},
	url = {https://doi.org/10.1088%2F1742-6596%2F1342%2F1%2F012094},
	year = 2020,
	month = {jan},
	publisher = {{IOP} Publishing},
	volume = {1342},
	pages = {012094},
	author = {C Buck and others},
	journal = {J. Phys (Conf. Ser.)}
}

@article{cogent,
  author = {Aalseth, C. E. and others},
  journal = {Phys. Rev. D},
  volume = {88},
  issue = {1},
  pages = {012002},
  numpages = {20},
  year = {2013},
  month = {Jul},
  publisher = {American Physical Society},
  doi = {10.1103/PhysRevD.88.012002},
  url = {https://link.aps.org/doi/10.1103/PhysRevD.88.012002}
}

@phdthesis{phil,
author={Barbeau, Phil},
school={University of Chicago},
year={2009},
}

@article{reactor_em_3, title={New bounds on neutrino electric millicharge from GEMMA experiment on neutrino magnetic moment}, volume={273–275}, rights={https://www.elsevier.com/tdm/userlicense/1.0/}, ISSN={24056014}, url={https://linkinghub.elsevier.com/retrieve/pii/S2405601415009633}, DOI={10.1016/j.nuclphysbps.2015.10.004}, abstractNote={Using the new limit on the neutrino anomalous magnetic moment recently obtained by GEMMA experiment we get an order-of-magnitude estimation for possible new direct upper bound on the neutrino electric millicharge | qν |∼ 1.5 × 10−12e0 (e0 is the absolute value of the electron charge) by comparing the neutrino magnetic moment and millicharge contributions to the total cross section at the electron recoil energy threshold of the experiment. This estimation is conﬁrmed by the performed analysis of the GEMMA data using established statistical procedures and a new direct bound on the neutrino millicharge absolute value | qν |< 2.7 × 10−12e0 (90%CL) is derived. This limit is more stringent than the previous one obtained from the TEXONO reactor experiment data that is included to the Review of Particle Properties 2012.}, journal={Nucl. Par. Phys. Proceedings}, author={Brudanin, Victor B. and Medvedev, Dmitry V. and Starostin, Alexander S. and Studenikin, Alexander I.}, year={2016}, month=apr, pages={2605–2608} }

@ARTICLE{gemma,
       author = {{Beda}, A.~G. and others},
        title = "{Gemma experiment: The results of neutrino magnetic moment search}",
      journal = {Phys. Par. Nucl. Lett.},
         year = {2013},
        month = {Mar},
       volume = {10},
       number = {2},
        pages = {139-143},
          doi = {10.1134/S1547477113020027},
       adsurl = {https://ui.adsabs.harvard.edu/abs/2013PPNL...10..139B}
}

@article{ve_orig, title={Neutrino electromagnetic form factors}, volume={39}, ISSN={0556-2821}, url={https://link.aps.org/doi/10.1103/PhysRevD.39.3378}, DOI={10.1103/PhysRevD.39.3378}, number={11}, journal={Phys. Rev. D}, author={Vogel, P. and Engel, J.}, year={1989}, month={Jun}, pages={3378–3383} }

@article{amsler,
title = "Temperature dependence of pure CsI: scintillation light yield and
decay time",
journal = "Nucl. Instr. Meth. A",
volume = "480",
number = "2",
pages = "494 - 500",
year = "2002",
issn = "0168-9002",
doi = "https://doi.org/10.1016/S0168-9002(01)01239-6",
url = "http://www.sciencedirect.com/science/article/pii/S0168900201012396",
author = "C Amsler and others",
}

@article{mos1,
author = {Moszynski, Marek and others},
year = {2005},
month = {01},
pages = {357-362},
title = {Energy resolution and non-proportionality of the light yield of
pure CsI at liquid nitrogen temperatures},
volume = {537},
journal = "Nucl. Instr. Meth. A",
doi = {10.1016/j.nima.2004.08.043}
}

@article{mos2,
title = "Application of large area avalanche photodiodes to study
scintillators at liquid nitrogen temperatures",
journal = "Nucl. Instr. Meth. A",
volume = "504",
number = "1",
pages = "307 - 312",
year = "2003",
issn = "0168-9002",
doi = "https://doi.org/10.1016/S0168-9002(03)00785-X",
url = "http://www.sciencedirect.com/science/article/pii/S016890020300785X",
author = "M. Moszynski and others"
}

@phdthesis{nadeau,
author={Nadeau, Patrick},
school={Queen's University},
year={2015},
}

@article{clark,
title = "Particle detection at cryogenic temperatures with undoped CsI",
journal = "Nucl. Instr. Meth. A",
volume = "901",
pages = "6 - 13",
year = "2018",
issn = "0168-9002",
doi = "https://doi.org/10.1016/j.nima.2018.05.066",
url = "http://www.sciencedirect.com/science/article/pii/S0168900218306855",
author = "M. Clark and P. Nadeau and S. Hills and C. Dujardin and P.C.F. Di
Stefano"
}

@article{liu,
	doi = {10.1088/1748-0221/11/10/p10003},
	url = {https://doi.org/10.1088%2F1748-0221%2F11%2F10%2Fp10003},
	year = 2016,
	month = {oct},
	publisher = {{IOP} Publishing},
	volume = {11},
	number = {10},
	pages = {P10003--P10003},
	author = {J. Liu and M. Yamashita and A.K. Soma},
	title = {Light yield of an undoped {CsI} crystal coupled directly to a
photomultiplier tube at 77 Kelvin},
	journal = {J. Instrum.}
}

@ARTICLE{woody,
author={C. L. {Woody} and others},
journal={IEEE Trans. Nucl. Sci.},
title={Readout techniques and radiation damage of undoped cesium iodide},
year={1990},
volume={37},
number={2},
pages={492-499},
doi={10.1109/23.106667},
ISSN={0018-9499},
month={April},}

@article{zhang,
	doi = {10.1007/s41605-018-0039-1},
	year = 2018,
	volume = {2},
	pages = {15},
	author = {Zhang, X. and others},
	journal = {Radiat. Detect. Technol. Methods },
}

@article{mik,
author = {Mikhailik, V. B. and Kapustyanyk, V. and Tsybulskyi, V. and Rudyk,
V. and Kraus, H.},
title = {Luminescence and scintillation properties of CsI: A potential
cryogenic scintillator},
journal = {Phys. Status Solidi (b)},
volume = {252},
number = {4},
pages = {804-810},
doi = {10.1002/pssb.201451464},
year = {2015}
}

@Article{nol,
author={Ponomarenko, Sergei A.
and others},
title={Nanostructured organosilicon luminophores and their application in highly efficient plastic scintillators},
journal={Nature Sci. Rep.},
year={2014},
month={Oct},
day={08},
publisher={The Author(s) SN  -},
volume={4},
pages={6549 EP  -},
url={https://doi.org/10.1038/srep06549}
}

@article{nol2,
author ="Starikova, T. Yu. and others",
title  ="A novel highly efficient nanostructured organosilicon luminophore
with unusually fast photoluminescence",
journal  ="J. Mater. Chem. C",
year  ="2016",
volume  ="4",
issue  ="21",
pages  ="4699-4708",
publisher  ="The Royal Society of Chemistry",
doi  ="10.1039/C6TC00979D",
url  ="http://dx.doi.org/10.1039/C6TC00979D",
}

@inproceedings{nol3,
author = {Sergey A. Ponomarenko and others},
title = {{Nanostructured organosilicon luminophores for efficient and fast
elementary particles photodetectors}},
volume = {10344},
booktitle = {Nanophotonic Materials XIV},
organization = {International Society for Optics and Photonics},
publisher = {SPIE},
pages = {49 - 58},
year = {2017},
doi = {10.1117/12.2273981},
URL = {https://doi.org/10.1117/12.2273981}
}

@inproceedings{nol4,
author = {Sergey A. Ponomarenko and Nikolay M. Surin and Oleg V. Borshchev
and Maxim S. Skorotetcky and Aziz M. Muzafarov},
title = {{Nanostructured organosilicon luminophores as a new concept of
nanomaterials for highly efficient down-conversion of light}},
volume = {9545},
booktitle = {Nanophotonic Materials XII},
organization = {International Society for Optics and Photonics},
publisher = {SPIE},
pages = {8 -- 16},
year = {2015},
doi = {10.1117/12.2187281},
URL = {https://doi.org/10.1117/12.2187281}
}

@Article{nol5,
author={O.V. Borshchev and N.M. Surin and M.S. Skorotetcky and S.A. Ponomarenko.},
title={High-efficient optical wavelength shifters: design, properties, application},
journal={INEOS OPEN},
year={2019},
publisher={The Author(s) SN  -},
volume={2 (4)},
pages={112},
doi = {10.32931/io1916r}
}

@INPROCEEDINGS{bigapd,
author={M. {McClish} and others},
booktitle={IEEE Nucl. Sci. Symp. Conf. Rec. 2004},
title={Characterization of very large silicon avalanche photodiodes},
year={2004},
number={},
pages={1270-1273 Vol. 2},
doi={10.1109/NSSMIC.2004.1462432},
ISSN={1082-3654},
month={Oct},}

@article{highQEapd, title={A study of low resistivity, deep diffused, silicon avalanche photodiodes coupled to a LaBr3:Ce scintillator}, volume={610}, ISSN={0168-9002}, url={https://www.sciencedirect.com/science/article/pii/S016890020901064X}, DOI={10.1016/j.nima.2009.05.072}, abstractNote={Radiation Monitoring Devices (RMD) has modified their production of deep diffused, planar silicon avalanche photodiodes (APDs), which resulted in significant performance improvements. This modification involves an alternative planar process to influence the p–n junction contour to create a planar bevel while using 4Ωcm n-type neutron transmutation-doped silicon wafers, where previously 30Ωcm silicon wafers were used. These new APDs still offer a high gain (∼103), but with an increased quantum efficiency and a reduced noise by a factor of 4–5, compared to our standard planar processed 30Ωcm APDs with the same detection area. We have characterized these new devices for their intrinsic and spectroscopic properties. In our study a 14×14mm2 APD, made from 4Ωcm silicon, was coupled to a 1cm3 LaBr3:Ce scintillator. We measured a FWHM energy resolution at 662keV to be 2.55% at room temperature (24°C).}, number={1}, journal={Nucl. Instr. Meth. A}, author={McClish, M. and Farrell, R. and Glodo, J. and Shah, K. S.}, year={2009}, month=oct, pages={207–209} }

@MastersThesis{jin,
    author     =     {Yifan Jin},
    school     =     {University of Tokyo},
    year     =     {2015},
    }

@article{belle,
title = "Study of a pure CsI crystal readout by APD for Belle II end cap ECL
upgrade",
journal = "Nucl. Instr. Meth. A",
volume = "824",
pages = "691 - 692",
year = "2016",
issn = "0168-9002",
doi = "https://doi.org/10.1016/j.nima.2015.07.034",
url = "http://www.sciencedirect.com/science/article/pii/S0168900215008633",
author = "Y. Jin and others"
}

\end{document}